\documentclass[%
 reprint,
 amsmath,amssymb,
 aps,
]{revtex4-1}

\usepackage{graphicx}
\usepackage{dcolumn}
\usepackage{bm}

\begin{document}

\preprint{APS/123-QED}

\title{Aggregation Dynamics of Active Rotating Particles in Dense Passive Media}

\author{Juan L. Aragones}
\thanks{These two authors contributed equally}
\author{Joshua P. Steimel}
\thanks{These two authors contributed equally}
\author{Alfredo Alexander-Katz}
 \email{aalexand@mit.edu}
\affiliation{
Department of Materials Science and Engineering, Massachusetts Institute of Technology, Cambridge, MA, 02139, USA.\\
}

\date{\today}

\begin{abstract}
Active matter systems are able to exhibit emergent non-equilibrium states due to activity-induced effective interactions
between the active particles. Here we study the aggregation and dynamical behavior of active rotating particles, spinners,
embedded in 2D passive colloidal monolayers, which constitutes one such non-equilibrium process. Using both experiments
and simulations we observe aggregation of active particles or spinners whose behavior resembles classical 2D coarsening.
The aggregation behavior and spinner attraction  depends on the mechanical properties of the passive monolayer and the
activity of spinners. Spinner aggregation only occurs when the passive monolayer behaves elastically and when the spinner
activity exceeds a minimum activity threshold. Interestingly for the spinner concentrations investigated here, the spinner 
concentration doesn't seem to change the dynamics of the aggregation behavior. There is also a characteristic cluster size
at which the dynamics of spinner aggregation is maximized as drag through the passive monolayer is minimized and the
stress applied on the passive medium is maximized. We also show that a ternary mixture of passive particles, co-rotating,
and counter-rotating spinners also aggregates into clusters of co and counter-rotating spinners respectively.
\end{abstract}

  

\maketitle

\section{Introduction}

Attractive interactions between particles in a homogeneous mixture induces the formation of clusters.
These clusters grow in time until the mixture separates into two distinct phases. The dynamics of phase
separation in binary mixtures is fairly well characterized~\cite{Shani}; and depends on the
dimensionality, thermodynamic conditions, and type of cluster growth. Often, these attractive interactions
are induced by direct chemical interactions. Alternatively, they can also be induced by electromagnetic,
phoretic~\cite{Palacci:2013eu}, collisions~\cite{Corte:2008iu} or hydrodynamic 
forces~\cite{Ishikawa:2008ht,Berke:2008cu,Drescher:2009cy}. Non-equilibrium interactions can also promote
particle aggregation. Because of the out-of-equilibrium nature of active matter systems, these are excellent
candidates to study novel mechanisms of particle aggregation and subsequent phase separation. 

Active matter systems are composed of active agents that consume energy from their environment and convert
it into motion or mechanical forces. The most prominent examples of these active systems are living organisms,
which exhibit striking emergent non-equilibrium behavior such as swarming, lining, vortexes, etc.~\cite{buhl_disorder_2006,darnton_dynamics_2010,ordemann_pattern_2003,wu_periodic_2009,zhang_collective_2010,
lin_dynamics_2014,toner_flocks_1998,ballerini_interaction_2008,cavagna_scale-free_2010,silverberg_collective_2013,szabo_phase_2006,hinz_motility_2014,wioland_confinement_2013,lin_dynamics_2014,sanchez_spontaneous_2012,joanny_biological_2010}. Synthetic active systems that are able to mimic and reproduce
some of the emergent behavior exhibited by living organisms can be used as model systems to study the underlying physical principles which govern
their behavior. The activity of the active components can convert energy locally into motion, as in living systems, or induced via an externally applied field or stimuli. Some examples of which include magnetic or electric fields, light-catalyzed chemical reactions, vibrating
granular beds and optical tweezers. Importantly, it is this activity which perpetually drives these systems out-of-equilibrium. However, most biological systems and processes are not composed of purely active components. In biological systems the
active or motile components, i.e. cells, are often surrounded by immobile, passive, or even abiotic interfaces. Investigating
emergent non-equilbrium behavior in such an artificial model system composed of active and passive components can 
potentially help distinguish what biological interactions can be attributed to purely physical phenomena and which interactions
require presumably physical and biological/biochemical stimuli.

\begin{figure}[h!]
\centering
\includegraphics[clip,scale=0.26,angle=0]{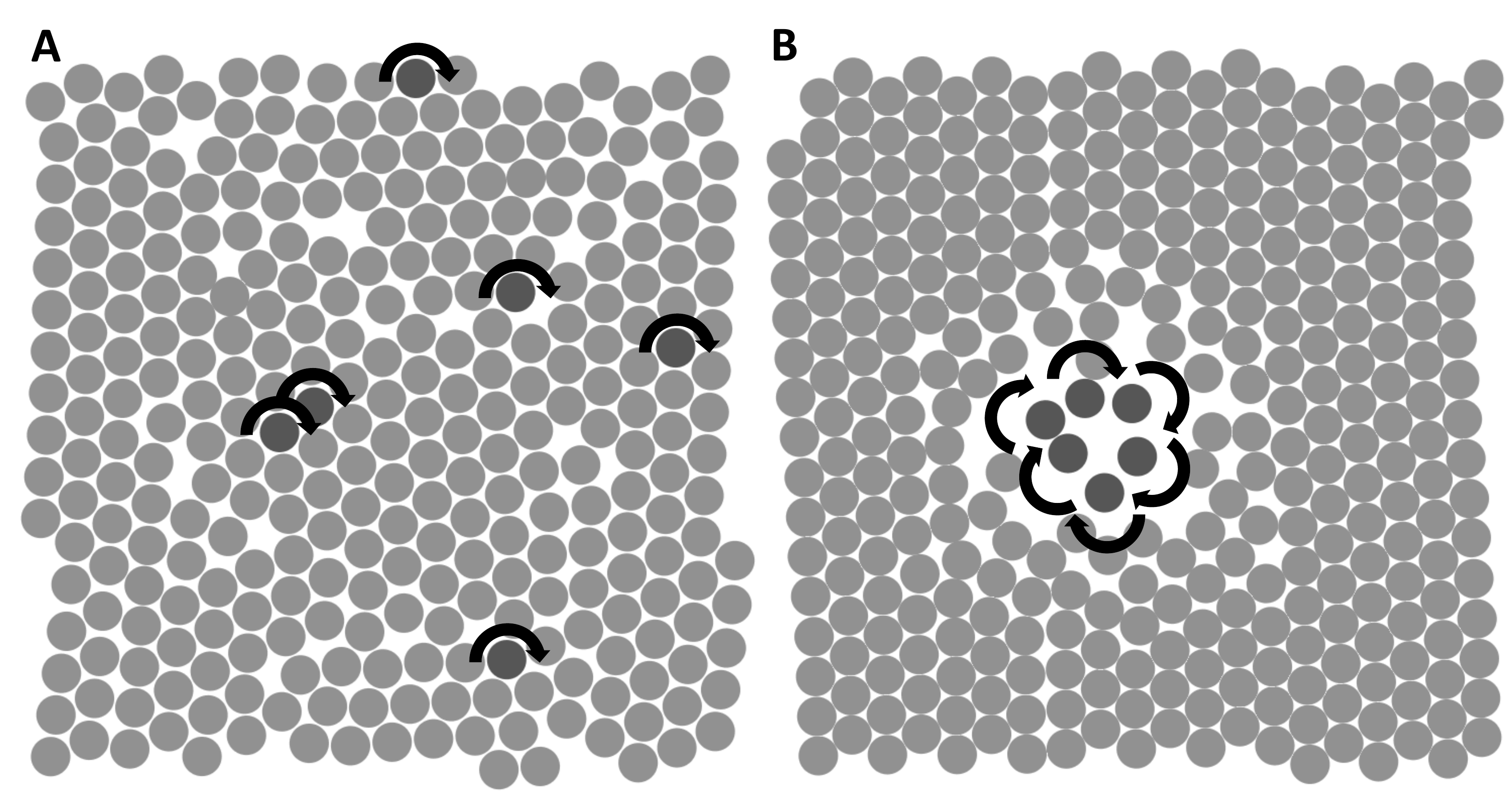}
\caption{
Schematic representation of the system. A) Co-rotating spinners randomly distributed within a monolayer
of passive particles of $\phi_A$ = 0.8, which under the action of the magnetic field rotate around the axis
perpendicular to the monolayer plane (i.e. z-axis). B) Spinner clusters form due to attraction between active
particles.  
}
\label{fig0}
\end{figure}

Here, we study a model active matter system that is composed of both passive and active components to study the 
aggregation of active particles. The active particles rotate in place, henceforth referred to as \textit{spinners}, and are 
embedded in a dense monolayer of passive particles, as schematically shown in Fig.~\ref{fig0}A. In this system, 
active particles exhibit a long-range attractive interaction~\cite{Aragones:2016cn,Steimel:2016jz}, which emerges
from the non-equilibrium nature of the system and it is mediated by the mechanical properties of the passive medium.
We observe that spinners embedded in a dense passive monolayer tend to aggregate forming clusters which grow
in time, as schematically shown in Fig.~\ref{fig0}. By means of experiments and numerical simulations we demonstrate
that this spinner aggregation is driven by the non-equilibrium attractive interaction induced by the mechanical properties
of the passive matrix. Moreover, we show that the spinners aggregation process follows a dynamics that resemble a 
coalescence process. The dynamics of the aggregation process depends on the mechanical properties of the monolayer
as well as the activity of the spinners. This type of non-equilibrium attractive interaction opens the door to controlling
the state of the system via control of the mechanical properties of the medium, activity of the spinners, and the density
of spinners.
  
\section{Materials and Methods}

Our synthetic model system is composed of active spinning particles and passive particles. The spinners are superparamagnetic
polymer-based magnetite particles purchased from Bangs Laboratories while the passive particles are composed of polystyrene purchased
from Phosphorex; both, active and passive, are 3 $\mu$m in diameter. We use a concentrated solution of spinners, $\approx 4 
\mu g/mL$, and passive particles, $\approx 0.8 mg/mL$, to study the aggregation and dynamical behavior of the spinners. 
The spinners are made active by externally applying a magnetic field which rotates around the axis perpendicular to the plane 
of the monolayer. The solution of spinners and passive particles is inserted into a channel (22mm (L) $\times$ 3mm
(W) $\times$ 300$\mu$m (H)), fabricated using a glass slide, spacer, and cover slip. Once the solution is inserted into the channel
it is sealed with epoxy and allowed to sediment for 10 minutes to form a dense monolayer, before being magnetically actuated.
The strength of the magnetic field is 5~mT, which is large enough to maintain alignment of the ferromagnetic particles with the
rotational frequency of the field. The magnetic field is actuated at an angular frequency, $\omega$, of 5~Hz for approximately
10 minutes. This rotational frequency corresponds to  Re = 1.25$\times$10$^{-6}$. We switch the magnetic field rotational sense
every 2 min.

In addition, we carry out numerical simulations of this system. In particular our coarse grained model consists of pseudo-hard
sphere particles~\cite{Jover:2012jy}, N=324, suspended on a fluid of density $\rho$ = 1 and kinetic viscosity $\nu$ = 1/6
modeled using the Lattice-Boltzman method. We use the fluctuating Lattice-Boltzmann equation~\cite{Dunweg:2008hb} with
$k_BT$ = 2$\times$10$^{-5}$ and the solver D3Q19. We discretized the simulation box in a three dimensional grid of 
$N_x \times N_y \times N_z = 101 \times 101  \times 20$ bounded in the z direction by no-slip walls and periodic boundary
conditions on the x and y directions. We set the grid spacing, $\Delta$x and time step, $\Delta$t, equal to unity. We apply the
bounce-back rule~\cite{Ladd:1994wb} to describe the interaction between the solid particles and the fluid. The particles are
treated as real solid objects~\cite{Ding:2003wl} of diameter $\sigma$ = 4$\Delta$x.  The particles settle on the bottom wall of
the channel forming a monolayer under the action of a gravitational force, F$_G$ = 0.005. The activity is achieved by imposing
a constant torque, which in general corresponds to Re = 0.72, unless otherwise noticed.

\section{Results and Discussion}

\begin{figure}[h!]
\centering
\includegraphics[clip,scale=0.35]{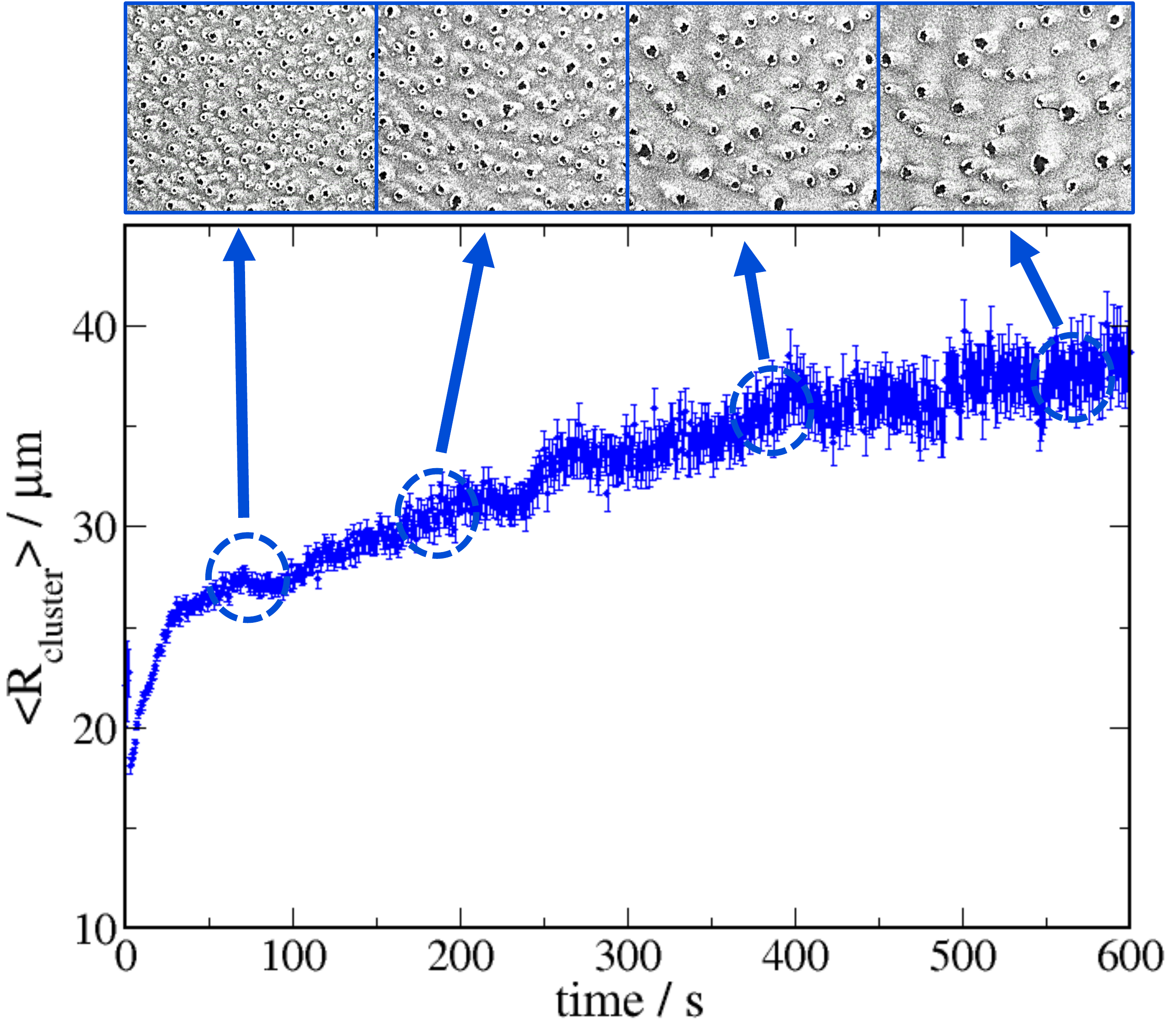}
\caption{Average spinner cluster radius, $<R_{cluster}>$, as a function of time. The top panel shows several experimental
snapshots, which clearly show the average size of the spinner clusters growing over time. The spinners correspond to the
darker spots.
}
\label{fig1}
\end{figure}

We study the behavior of active rotating particles within monolayers of passive particles.
We prepare a dense monolayer composed of passive polystyrene particles, particle area fraction
$\phi_A \approx 0.7$, and doped it with a small particle fraction, about 0.5\%, of active superparamagnetic particles.
Upon actuation of the magnetic field, we observe that spinners aggregate forming nearly circular actively rotating
clusters, whose average radius, $<R_{cluster}>$, grows with time, as shown in Fig.~\ref{fig1}.
These clusters of spinners can be seen growing over time in the experimental snapshots at the top of the figure,
where the spinners are the darker spots in the snapshots. This behavior is different than that observed in a system
of purely active ferromagnetic and superparamagnetic particles~\cite{Kalontarov:2010go,Promislow:1995ky,Richardi:2011ev,Gao:2012bx}. At similar particle are fractions and frequencies in the purley active system the aggregation of particles is dominated by the magnetic dipole-dipole interaction. We observe the
formation of small chains or random aggregates~\cite{Richardi:2011ev}, which size grows slightly with time.
Therefore, the passive monolayer is not only modifying the shape of the aggregates, but also increases the
range of the interaction. In fact, in pure equilibrium arguments one would expect the monolayer to strongly retard
the aggregation because of the high viscosity of the monolayer.    

\begin{figure}[h!]
\centering
\includegraphics[clip,scale=0.4]{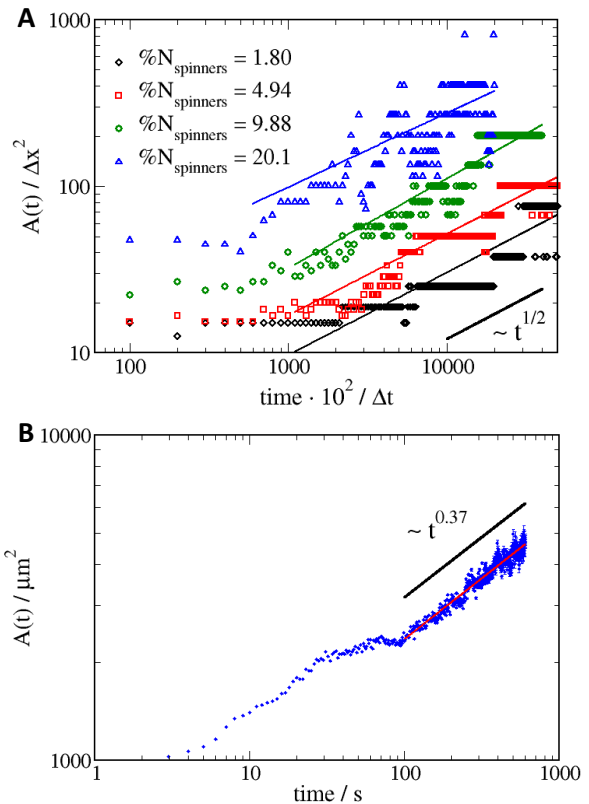}
\caption{A) Log - log scale for the spinner cluster size (area), A(t), as a function of time in a passive monolayer of 
$\phi_A$ = 0.8 for four different spinner concentrations: 1.8\% (black circles), 4.94\% (red squares),
9.88\% (green diamonds) and 20.1\% (blue triangles) in simulations.  
B) Log - log scale for the spinner cluster length from experiments as a function of time in a passive monolayer of
$\phi_A \approx $ 0.7.
}
\label{fig2}
\end{figure}

We carry numerical simulations of this system to study the role of the passive matrix in the spinner aggregation
process, and consider whether magnetic dipole-dipole interactions play a mayor role in the spinner aggregation process,
particularly at large cluster sizes. Therefore, our coarse grained
model neglects the dipole-dipole interactions to isolate the effect of the passive matrix in the spinner aggregation
process. In agreement with the experimental results we observe that spinners aggregate forming circular
clusters when embedded in dense monolayers of passive particles of $\phi_A$ = 0.8.
We calculate the time evolution of the area of the active clusters, A(t), for the experimental and simulation trajectories,
as shown in Fig.~\ref{fig2}. We again observe that the size of the clusters grows with time. Hence, the presence of the
passive monolayer promotes the aggregation of the spinners, even in the absence of magnetic dipole-dipole interactions.
In addition, in the simulations we observe that the time evolution of the domain length, A(t), presents important fluctuations
due to the absence of dipole-dipole interactions. When two clusters collide, the clusters split in pieces and
while the new merged cluster is re-configuring the size of the clusters fluctuates, as shown in Movie 1. 

We have recently shown that an attractive interaction emerge between two co-rotating particles, or spinners,
if embedded in dense passive monolayers~\cite{Aragones:2016cn,Steimel:2016jz}. This emergent attractive
interaction and the subsequent non-equilibrium phase separation is thus mediated by the elasticity of the medium
and the ability of the spinners to stress that medium. Under the actuation of the rotating magnetic field, the spinners
rotate around the axis perpendicular to the substrate generating a rotational fluid flow~\cite{Climent:2007du,2001PhRvE..64a1603G,Grzybowski:tt}. This causes the surrounding passive particles to rotate due to the momentum transferred
through the fluid in which the particles are suspended. In addition, at small but finite Re, the spinner's rotational motion
produces a so-called secondary flow due to the fluid inertia, which pushes away the nearest shell of passive particles,
effectively compressing the passive monolayer. Thus, two co-rotating spinners apply compressive and shear stresses
on the passive particles located in between the spinners, referred to as the bridge. This produces a stochastic, but steady
degradation of the bridge, which allows the spinners to approach resulting in an attractive interaction~\cite{Aragones:2016cn}. 
Moreover, depending on the mechanical properties of the passive monolayer, this attractive interaction between active
rotating particles may be of a very long-range nature~\cite{Steimel:2016jz}.  

\begin{figure}[h!]
\centering
\includegraphics[clip,scale=0.4]{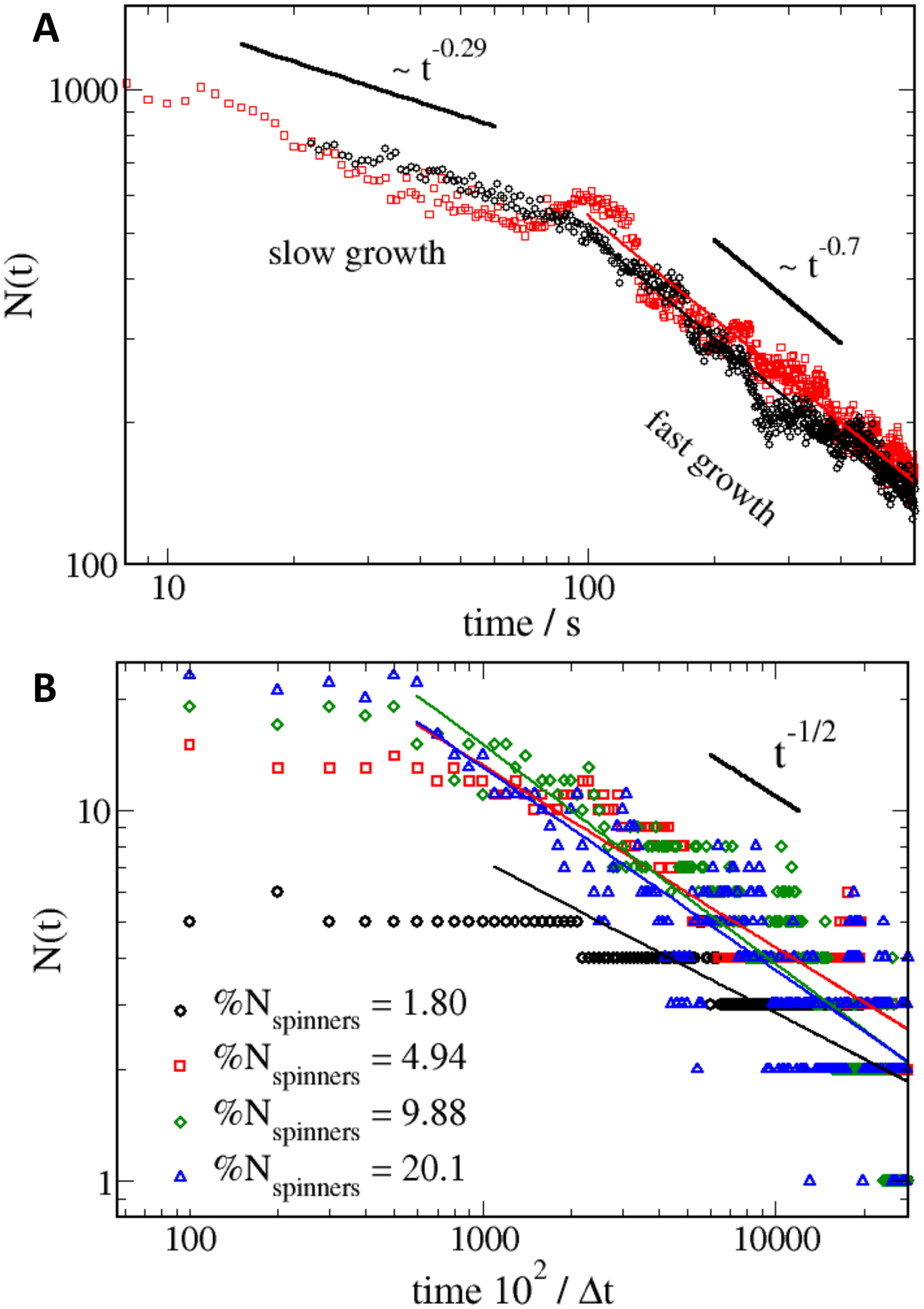}
\caption{A) Log - log scale for the number of clusters as a function of time as obtained from the
experiments at $\phi_A \approx$ 0.7. The two sets of data corresponds to two independent experiments.
B) Log - log scale for the number of clusters as a function of time as obtained from the
simulation model for four different spinner concentrations at $\phi_A$ = 0.8: 1.8\% (black circles),
4.94\% (red squares), 9.88\% (green diamonds) and 20.1\% (blue triangles).
}
\label{fig3}
\end{figure}

The solid-like character of the passive monolayer induces an attractive interaction between the active particles resulting
in the aggregation of the spinners embedded in passive matrixes. However, we do not observe the complete phase
separation of the system into passive and active domains for the actuation time period investigated. Instead, spinners
aggregate forming clusters which grow with time embedded within the passive matrix. The dynamics of this aggregation
process resembles a spinodal decomposition process, in which active clusters coalesce. From the experimental trajectories
we compute the time evolution of the number of active clusters, N(t), as shown in Fig.~\ref{fig3}A. We observe two different
dynamical regimes within the experimental time scale. At short time scales (less than 100s) the clusters exhibit an initial
regime of slow cluster growth, or an almost constant number of clusters. This is followed by another regime (after 100s)
where the spinner aggregate at a much faster rate. As it can be seen in Fig.~\ref{fig3}A, the scaling of the number
of clusters with time in this regime is characterized by an exponent of $\approx$ -0.7. We also analyze the dynamic
scaling of the aggregation of spinners in our simulation model. In this case, we also observe two dynamical regimes:
i) An initial slow decrease in the number of clusters (i.e. growth of the cluster sizes), followed by a regime with a dynamic
scaling of exponent $\approx$ -0.5, as shown in Fig.~\ref{fig3}B. Moreover, this dynamic scaling seems independent
on the spinner concentration. The $t^{1/2}$ dynamical scaling of the cluster growth has been observed in simulations
conducted by Vicsek and several distinct purely active systems, although  the origin of this scaling behavior is still unclear~\cite{Dey:2012kb,2013PhRvL.110e5701R}. However, we believe that the origin of this scaling behavior is similar to that
observed in traditional 2D coarsening ~\cite{1974PhRvL..33.1006B}.   

We hypothesize that the difference between the exponents of the dynamic scaling observed in experiments and
simulations is due to the magnetic interaction between clusters of spinners, which increases the strength of the
spinner-spinner attraction at short distances. Additionaly, this also helps to stabilize spinner clusters.
If we also assume a dynamic scaling factor for the slower initial regime in the spinner aggregation process, we
obtain for experiments and simulations exponents, $\approx$ 0.29(2) and $\approx$ 0.04(5), respectively. In this
first aggregation regime the differences between the experimental and simulation dynamic scaling exponents are
much greater than for the second regime. This behavior is in agreement with our hypothesis that the effect of the 
magnetic dipole-dipole interactions significantly increases the spinner aggregation dynamics. At the beginning of
the aggregation process, the active clusters are small and the dipoles of the particles are more 
easily aligned, which results in higher magnetization of the clusters. In contrast, as the size of the clusters increases,
some of the dipoles of the particles are frustrated by the cluster structure, which results in a smaller 
magnetization of the clusters as their size increases. Therefore, the magnetic interaction is more relevant between
smaller clusters than between bigger clusters.
   
\begin{figure}[h!]
\centering
\includegraphics[clip,scale=0.4]{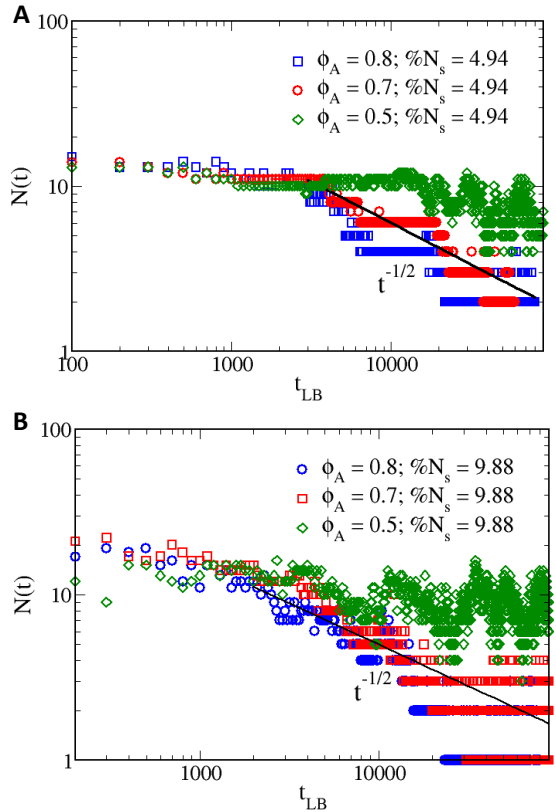}
\caption{A) Log - log scale of the time evolution of the number of active clusters at Re = 0.72 and
a spinner concentration of 4.94\% within passive monolayers of $\phi_A$ = 0.8 (blue squares), 0.7
(red circles) and 0.5 (green diamonds). 
B) Log - log scale of the time evolution of the number of active clusters at Re = 0.72 and
a spinner concentration of 9.88\% within passive monolayers of $\phi_A$ = 0.8 (blue squares), 0.7
(red circles) and 0.5 (green diamonds). 
}
\label{fig4}
\end{figure}   
   
The mechanical properties of the passive media determines the interaction between the spinners and thus, the
dynamics of the spinner aggregation. The mechanical properties of the monolayer can be calculated by measuring the
mean square displacement (MSD) of the particles in the monolayer in the absence of active particles, specifically the storage and loss modulus, G' and G'' respectively~\cite{Mason:1995um,
Mason:2000gm}. In simulations, we observe that monolayers of hard-sphere particles at area fractions $\phi_A > 0.7$
respond as viscoelastic materials, behaving as a viscous system at low frequencies and as a solid-like material at high
frequencies~\cite{Aragones:2016cn}. However, for $\phi_A < 0.7$ the monolayer behaves as a viscous material over the
entire frequency range. To study the effect of the mechanical properties of the monolayer on the dynamics of the spinner
aggregation, we investigate, by means of our simulation model, spinners embedded in passive monolayers at different
particle area fractions $\phi_A$ = 0.5, 0.7, and 0.8. As it can be seen in Fig.~\ref{fig4}, spinners in monolayers of particle area
fraction of $\phi_A$ = 0.8 and 0.7 follow similar scaling laws. However, at an area fraction of $\phi_A$ = 0.5, the spinners
do not aggregate within the simulation time scale, as shown by the green diamonds in Fig.~\ref{fig4}. The small amount
of spinner aggregation observed in Fig.~\ref{fig4} is due to spinners being initially positioned together or close enough so
that the removal of a single passive row of particles was required. It should also be noted that for a more dilute concentration
of spinners no aggregation is observed in the simulation timescale (data not shown). In addition, at these particle area
fractions the monolayer is unable to maintain spinners within a cluster and thus, the number of clusters exhibit large
 fluctuations. Thus, the presence of a passive monolayer that behaves as a solid-like material induces an attractive interaction
between the active rotating particles, which results in aggregation of spinners. Interestingly, the dynamics of the spinner
aggregation seems to be independent of the storage modulus, G', which is higher for monolayer of $\phi_A$ = 0.8 than for 
monolayer of $\phi_A$ = 0.7. This might be due to canceling of two competing effects. On one hand, the effective
interaction grows with G', but on the other hand the motion of the medium is controlled by $\eta$, which also grows. 

\begin{figure}[h!]
\centering
\includegraphics[clip,scale=0.35]{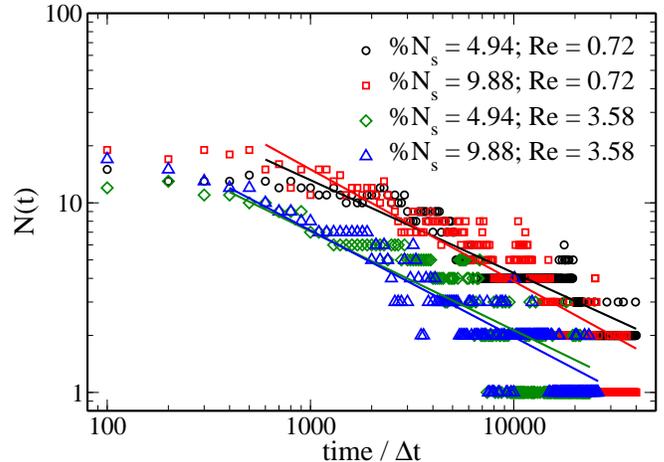}
\caption{Aggregation dynamics of spinners within passive monolayers of $\phi_A$ = 0.8 at two different rotational
frequencies Re = 0.72 (black and red symbols) and 3.58 (blue and green symbols) and spinner concentrations 
4.94\% (circles and diamonds) and 9.88\% (squares and triangles).}
\label{fig5}
\end{figure}

For spinners in a passive monolayer with a packing fraction $\phi_A > 0.7$ the mechanics of the passive monolayer also
plays an important role in keeping the cluster of active particles together. We have previously reported that for a system
composed of purely active particles the spinners will repel due to the secondary flows generated by the spinners
~\cite{Aragones:2016cn,Goto:1hu}. The spinners within the cluster should then repel, but the passive monolayer exerts
a force on the spinners that serves to stabilize the active cluster. This is evident from the stochastic fluctuations shown
in the cluster size time evolution, as shown in Fig.~\ref{fig2}B and \ref{fig4}, which corresponds to clusters breaking and
reforming during the aggregation process. The size of the fluctuations increases with the spinners' concentration, due to
the bigger size of the clusters.

Aside from the mechanical properties of the monolayer, the other requisite for spinner attraction is the ability to
stress the passive monolayer. Therefore, we also explore the effect of the spinners' activity on the aggregation 
dynamics by applying different rotational frequencies, Re = 0.1, 0.72 and 3.58, to spinners embedded in passive
matrixes of $\phi_A$ = 0.8. In agreement with our previous observations for the spinner-spinner interaction in
passive environments~\cite{Aragones:2016cn,Steimel:2016jz}, we observe there exists a minimum threshold of
loading stress, or spinner activity, for the spinner attractive interaction to be important. Spinners rotating at Re smaller
than 0.1 do not aggregate. At these activities the stress applied to the passive monolayer is not large enough to
promote the occurrence of yielding events, which ultimately results in spinner aggregation~\cite{Aragones:2016cn}. 
On the contrary, spinners rotating at Re $\ge$ 0.72 do aggregate, and the higher the rotational frequency, the faster
the evolution of the system, as shown in Fig.~\ref{fig5}. Interestingly, the dynamic scaling exponent of the spinner
aggregation seems to be independent on the rotational frequency. However, the range of the initial dynamical regime,
which probably corresponds to the fastest growing unstable composition mode, shifts towards shorter times.
The spinner-spinner attraction in passive matrixes follows activated dynamics~\cite{Steimel:2016jz}. The monolayer
region in between the two spinners (i.e. the bridge) needs to be loaded before it yields, which has an associated
time scale. This time scale depends on the mechanical properties and configuration of the monolayer. 
If the stress applied by the spinners overcomes this time scale, the passive particle mobility increases, resulting
in yielding events~\cite{Aragones:2016cn}, which leads to the erosion or degradation of the bridge. Therefore,
the higher the rotational frequency of the spinners (i.e. Re), the shorter time required to stress the bridge
and thus, as the  frequency of the spinners increases the faster the growth of the clusters at short time scales.  
The differences observed between the experiment and simulations on the Re comes from the approximations
made in our simulation model. For example, in our simulations the momentum transfer between the spinner and
neighboring particles comes exclusively from the fluid, while in the experiment friction and collision between
particles may play an important role in transferring momentum.        

\begin{figure}
\centering
\includegraphics[clip,scale=0.4]{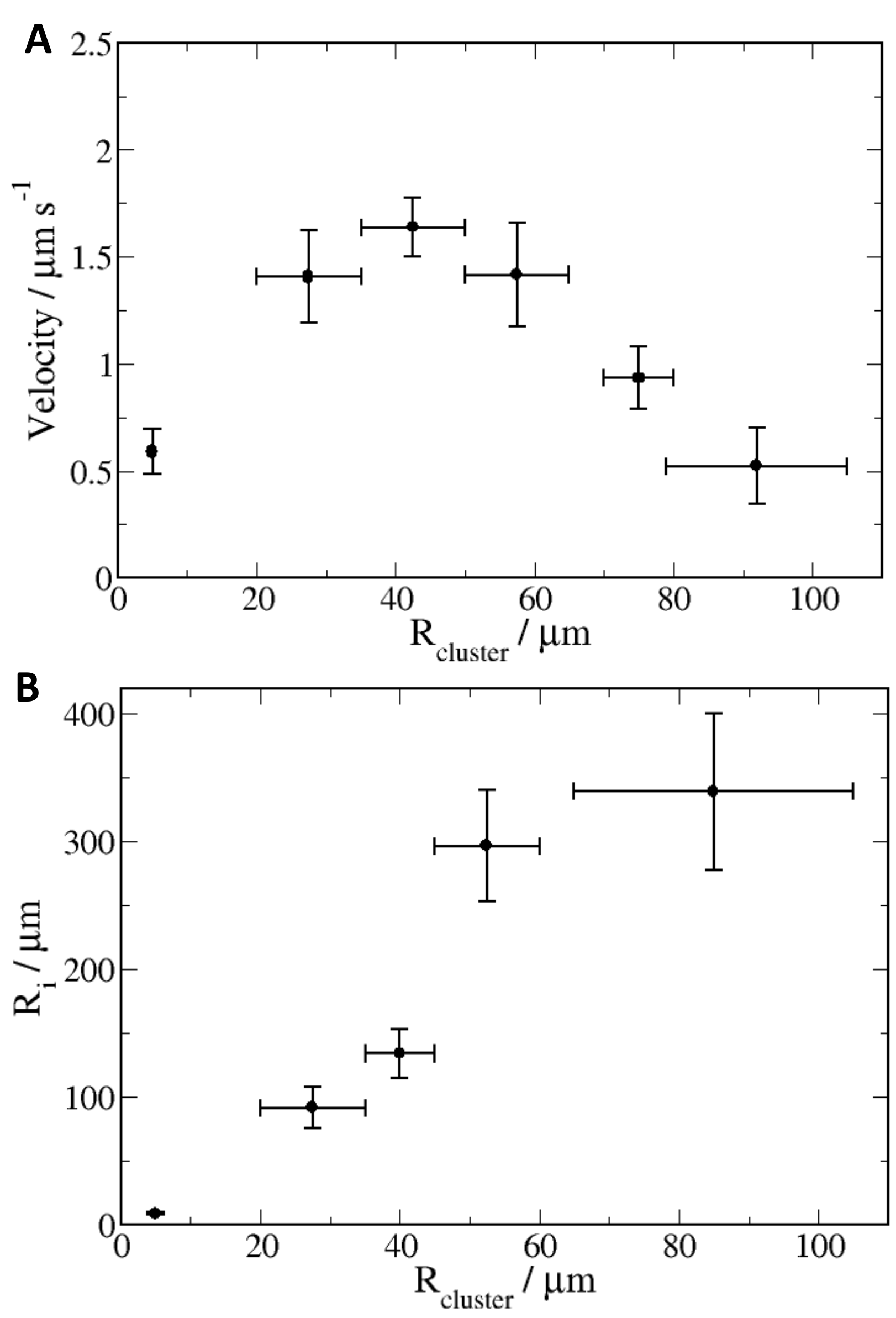}
\caption{A) Average approaching velocity of the active clusters as a function of the cluster size (i.e. radius of the cluster).
B) Range of the attractive interaction between active clusters as a function of the cluster size. Error bars corresponds
to the standard deviation of the trials.}
\label{fig6}
\end{figure}

We further investigate the microscopic details of the spinner aggregation process by tracking the active clusters
over time, noting when clusters collide, initial separation distances between clusters which merge, and the velocity
at which clusters approach, as shown in Fig.~\ref{fig6}. In Fig.~\ref{fig6}A, the velocity at which spinner clusters
approach as a function of cluster size is presented. We observe that there is a maximum velocity associated
with a cluster size of approximately 45$\mu$m. This behavior of cluster velocity as a function of size reveals two
competing effects involved in the mobility of the clusters, and therefore their aggregation. The effect which opposes
spinner aggregation is the effective drag, which opposes the movement of the clusters through the monolayer and the
drag increases with cluster size. Meanwhile the stress that the spinner cluster can exert on the monolayer increases
with the size of the cluster. This increases the frequency of the yielding events resulting in the degradation of the bridge
and spinner aggregation.
In addition, we observe that the range of the attractive interaction between clusters, R$_i$, increases with the cluster size,
as shown in Fig.~\ref{fig6}B. The individual spinner clusters were tracked and as the clusters collide and form bigger clusters
the initial distance between colliding clusters was calculated and plotted as a function of the average of the two colliding
clusters. As discussed above, the stress exerted on the monolayer increases with the cluster size.  Therefore, this increase
of the stress on the monolayer produces higher mobility of the passive particles of the monolayer, which results in longer
ranged interactions.

\begin{figure}[h!]
\centering
\includegraphics[clip,scale=0.325]{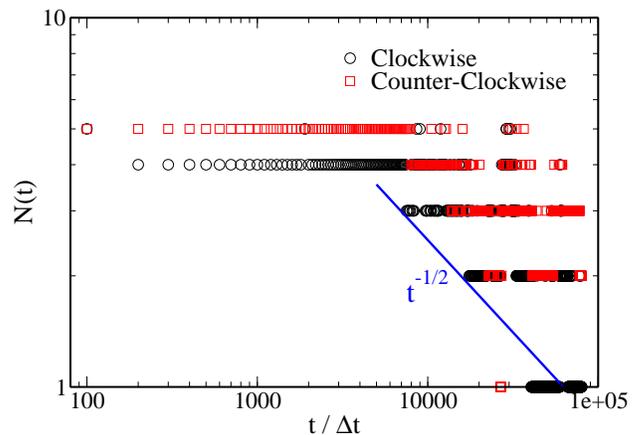}
\caption{Aggregation dynamics of spinners rotating clockwise (black circles) at a concentration of 1.54\%, and
clockwise (red squares) at a concentration of 1.54\% within a passive monolayers of $\phi_A$ = 0.8.}
\label{fig7}
\end{figure}

Finally, we explore the behavior of a ternary mixture in which a passive monolayer is doped with a symmetric mixture
of spinners rotating in opposite senses, clockwise and counter-clockwise. In our previous work, we demonstrated that
while two co-rotating spinners embedded in a passive matrix experience an attractive interaction, counter-rotating
spinners exhibit a repulsive interaction in dense passive environments~\cite{Aragones:2016cn}. We observe that
spinners in dense passive monolayers tend form clusters of co-rotating particles and thus, we observe the formation
of three different {\it phases}: i) passive particles, ii) spinners rotating clockwise and iii) spinners rotating counter-clockwise,
as shown in Fig.~\ref{fig7}. This is the result of the attractive interaction between spinners rotating on the same direction,
and the repulsive interaction between spinners rotating on opposite directions. 

\section{Conclusions}

We studied the aggregation of active rotating particles embedded in passive monolayer. We demonstrate that the
non-equilibrium attractive interaction between spinners within dense passive matrixes~\cite{Aragones:2016cn,Steimel:2016jz}
results in their aggregation. This aggregation resembles a 2D coarsening~\cite{1974PhRvL..33.1006B}, which
has also been described for other pure active systems~\cite{Dey:2012kb,2013PhRvL.110e5701R}. Although, the
system size we can reach does not allow us to unambiguously determine the dynamic scaling exponent of the spinner
aggregation process. We explore the effect of the particle area fraction of the monolayer, spinner concentration and
spinner activity on the aggregation behavior. We observe that the monolayer must behaves as a solid, $\phi_A >$ 0.7,
in order to observe spinner aggregation. In addition, for the spinners to stress the monolayer and thus, produces yielding
events that result into the attraction of spinners, there is a minimum activity threshold, Re $>$ 0.1 in simulations.
Interestingly, the aggregation dynamics seems to be independent of the spinner concentration. We also study the microscopic
details of the cluster aggregation. We observe that spinner clusters move faster as the size increases up to reach a velocity
maximum at around $R_{cluster}$ = 45~$\mu$m. Finally, we show that a ternary mixture of passive particles, co-rotating and 
counter-rotating spinners results into the formation of cluster of spinner with the same sense of rotation. This is due
to the fact that co-rotating spinners with in a dense passive monolayer attract each other while counter-rotating
spinner repel.  

\section{Acknowledgments}

This work was supported by Department of Energy BES award \#ER46919 (theoretical and simulation work)
and the Chang Family (experimental work). 

%

\end{document}